# INVESTIGATING THE AWARENESS OF APPLYING THE IMPORTANT WEB APPLICATION DEVELOPMENT AND MEASUREMENT PRACTICES IN SMALL SOFTWARE FIRMS

Faudziah Ahmad[1], Fauziah Baharom[2] and Moath Husni[3]

[1,2,3]College of Arts and Sciences, Universiti Utara Malaysia (UUM)
[1] `fudz@uum.edu.my`, [2] `fauziah@uum.edu.my`, [3] `tarawneh80@yahoo.com`

## ABSTRACT

*This paper aims to discuss the pilot study and analysis of the current development and measurement practices in Jordanian small software firms. It is conducted because most developers build web applications without using any specific development method and don't know how to integrate the suitable measurements inside the process to improve and reduce defect, time and rework of the development life cycle. Furthermore the objectives of this pilot study are firstly; determine the real characteristics of small software firms in Jordan. Secondly, investigate the current development and measurement practices. Thirdly, examine the need of new development methodology for building web application in small software firms. Consequently, Pilot survey was conducted in Jordanian small software firms. Descriptive statistics analysis was used to rank the development and measurements methods according to their importance. This paper presents the data, analysis and finding based on pilot survey. These actual findings of this survey will contribute to build new methodology for developing web applications in small software firms taking to account how to integrate the suitable measurement program to the whole development process and also will provide useful information to those who are doing research in the same area.*

## 1. INTRODUCTION

The use of Web-based applications which ranges from simple to sophisticated applications has become popular in our daily life as results of the rapid growth internet technology and environment[1]. Web application is defined as a "Web system which consists of Web server, network, HTTP and browser, in which user input (navigation and data input) affects the state of the business" [2]. In general Web-based applications differ from other traditional applications in term of high reliability, high usability, security, more technologies, shorter time to market, shorter product life cycles and continuous maintenance [3].

These applications need to be developed in a systematic way in order to achieve the software development goals such as delivered on time, within budget and with expected requirements. The systematic way of software development can be obtained through the use of an appropriate methodology. A methodology is a comprehensive, multiple-step approach to system development that guides developers to clearly understand the development process and influences the quality of the final product; it describes both the activities to be carried out and the deliverables that should be produced at the end of each activity [4]. Furthermore, it gives a full set of concepts and models which are internally self consistent and provides a collection of rules and guidelines [4]. Software companies which are involved with developing Web applications must follow a specific methodology to produce a high quality final product. Currently 85% of software companies are categorized as small software firms [5]. Small software firms are referring to any organization or company that has approximately 10 to 50 employees [6] [7] [8]. Small software firms are a very important sector in many countries such as US, Canada, China, India, Finland and Ireland as they provide substantial growth to these countries economy [5].

Unfortunately this particular type of organizations face a lot of problems such as project management, staffing, inappropriate process and methods, lack of risk management, lack of project control, limited resources for business development, limited staff skills and limited Quality Assurance adoption [7][9][10]. These problems motivate researcher to find ways for improvement. On the other hand the current development methods which used by small software firms still not aware for applying the important development and measurement practices [11][12][13]. Software measurement is defined as understand, control, predict and improve software development project which is useful for reduce defects, rework and cycle time [11][13].based on the definition, measurement must be integrated to the whole process not applying just on a specific stages of development process.

In Jordan, most of software firms at large are considered as small firms and government of Jordan has little knowledge on the quality of services or products produced by small firms in Jordan [12]. Therefore, an empirical study using survey technique was conducted in Jordan to investigate Web application development practices in small software firms.

This paper presents the findings on the pilot survey conducted in Jordan. Results of the survey indicates that there is a need of a new methodology for small software firms to follow and adopt in order to get a high quality product with in time and budget constraints. Consequently, the aim of this study to investigate the development and measurement practices for developing web applications in small software firms. This paper is categorized into four sections introduction, methodology, findings and conclusion.

## 2. RELATED WORK

The high changing requirement environment and shorter product life cycles and continuous maintenance makes the Web application development is very important unique sector in the software engineering [3]. As results many development methodologies have been proposed to address these unique challenge and characteristics of web applications. One of these methodologies is the conventional development methodologies like waterfall and spiral. However, these methodologies not adequate for Web applications development because they are not built mainly for developing Web applications and cannot address web applications unique characteristics with high changing requirement environment [14][15].

Agile development methodologies is proposed to solve the problem that faced by using the conventional methodologies in developing any software, where there is specific agile methodologies are used for developing software in small teams and projects, the most popular agile methodologies to be used for small software firms is Extreme Programming (XP) and SCRUM [16] [17][18]. However, XP has poor documentation, lack of management practices and it also does not handle requirement traceability and subsequent changes in requirements [18][19][20]. On the other hand Scrum has a lack of development practices because it does not define any specific software development techniques for the design and implementation phase and it has nothing to do about testing to ensure the quality of product [18][20].

International Journal of Computer Science & Information Technology (IJCSIT) Vol 3, No 6, Dec 2011Both XP and Scrum does not have any measurement program for mentoring the process [18][20]. Based on the above discussion there is a need for new web applications development methodology for small software firms integrating with suitable measurement program for monitoring the process.

There are numerous previous empirical studies conducted for discuss and address the software process best practices, the authors of these researches notify and advice the software developers to dominant, prevalent, common, best practices while they develop there software. However there is a lack of research to date to determine the actual current use of these practices [21]. A survey differentiates between the practices used by European firms and the practice that used by the Japanese firms [22], but a study conducted on the software management practices in US, Japan and Western Europe firms [23], concludes that companies in these countries used the same. Other researchers have focused on a particular location, for example, using the system development methodologies in Malaysia [24].

The most recent and related to the work that has been done in this paper are [12][21][25][26]. These studies indicate that there is alack of awareness of deploying the important development practices during the process in the targeted organizations.

Table 1 describes and summaries some of the recent empirical studies related with Web applications development and small software firms practices.

Table 1. Recent studies on web and sofware practives

| Study name | Respondent and data collection method | Objectives |
|---|---|---|
| A Survey of Web Engineering in Practice [25] | The respondent of this survey are the Web developers, and the data collection method was interview. | - To identify the major issues facing the development of Web based systems.<br><br>- To determine which, if any, traditional software engineering practices and techniques were being successfully applied. |
| An Evaluation Of Software Development Practice And Assessment-Based Process Improvement In Small Software Development Firms [21] | The respondent of this study are the software developer and managers inside small software firms and methods used for collect the data is questionnaires. | - To provide a much better understanding of practices used by small software development firms.<br><br>- To encourage these firms to adopt the best practice for improving the quality of the processes in use. |
| A Survey of Web Engineering Practice in Small Jordanian Web Development Firms [12] | The respondent of this study are the Web developer inside small software firms and the data collection method is questionnaires. | The goal of this survey is to show the level of Web engineering best practices adoption in the Jordanian small software firms. |

149

44

| A Survey on the Current Practices of Software Development Process in Malaysia [26] | The respondent of this survey are the Mangers, technical directors and developers on the Malaysian software companies and the data collection method is questionnaire. | - To determine the deployment of software development life cycle models.<br>- To determine the awareness of user involvement during the process and improving the developers skills.<br>- To identify quality problems and the extent of software reuse. |
|---|---|---|

## 3. METHODOLOGY

Survey approach was selected to be used for conducting the pilot study. A self-completion questionnaire is used as an instrument for collecting data and it was developed based on the literatures of web applications and software development. . The pilot study was conducted to check the reliability and validity of the questionnaire and to enhance the instruments and procedures.

The survey was conducted into three main stages: questionnaire design and formulation, data collection and data analysis.

### 3.1. Questionnaire Design and formulation

Questionnaire was adopted to be the data collection instrument for this survey. Therefore, this instrument developed and formulate based on literatures from web applications development and software development previous studies such as [12][25][26]. The questionnaire design consisted of three main sections: demographic information, development and measurement issues and web application development practices. Furthermore, the questionnaire sections included forty three questions and used open-end and closed-end questions. Mail questionnaire and interviews were used as the instruments for gathering and collecting data. However, this paper discussion concentrates only on the first two parts.

### 3.2. Data Collection

Conducting the pilot or pre-test a survey give the researcher good assistance before performing a full empirical study [27]. Firstly, conducting the pilot study allows the researcher to classify the types of responses for each question. Secondly, it provide as a quality assurance for grammar, sentence structure, and clarity. Lastly, a pilot survey is considered as an additional measure to maximize the effectiveness of a survey. The pilot survey should be directed to small group of respondents who are a similar as possible to the population of study.

In this pilot study, twenty three small software firms had been selected randomly and the study was conducted through questionnaire whereas the respondents were developers and mangers of small software firms. The questionnaire had already been formulated and prepared to be tested. One questionnaire was given to each respondent who answered the questions with the researcher guidance. The time required for answering the questionnaire was measured and any difficulties on answering the questions were discussed. The pilot survey has determined that respondents were able to answer the questions listed in the questionnaire. Pilot respondents advised for minor modifications on some items in the questionnaire and therefore prior to the actual survey, the feedbacks were used to refine the actual questionnaire.

## 3.3. Data Analysis

The data were coded and entered in SPSS version 14.0 (Statistical Package for Social Science) for analysis. Frequency and percentage were used to categories the demographic data variables. Cross tabulation and multi response techniques were used to calculate the results for development and measurement issues part.

## 4. FINDINGS

### 4.1. Demographic Data

#### 4.1.1. Company Size

This section clarifies the number of employee of each company in the pilot study. The majority of respondents indicate that their companies have 10-30 employees (52%) followed by 31-50 employees (44%) and only (4%) of companies have less than 10 employees. See Figure 1.

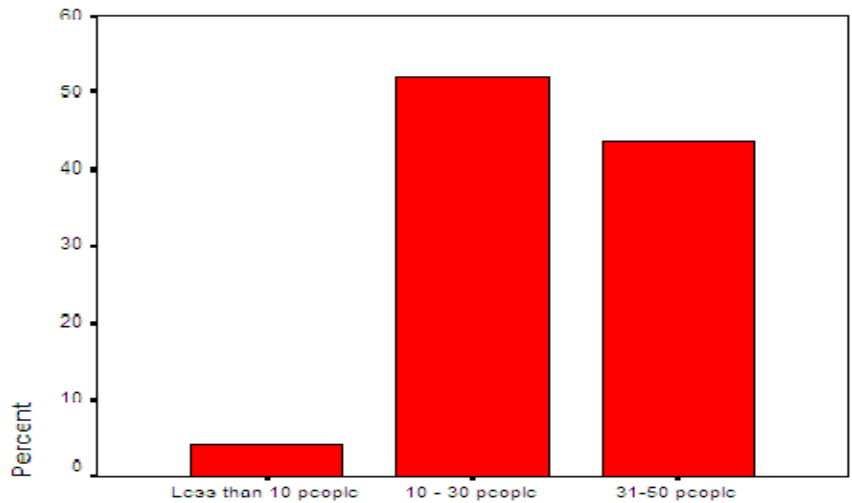

Figure 1. Company size

#### 4.1.2. Position and Experience

In this section respondents were asked about their position, experience. Table 2 demonstrates the distribution of respondent's position and the experience of years working in their companies. The data was then analyzed using cross tabulation analysis. The results obtained form the analysis showed that (52%) of respondents are 3-10 years of experience and most of them are team leaders (22%) followed by software engineering process group member (17%), technical members are (9%) and managers are (4%). On the other hand, 48%of respondents are less than three years of experience the majority of them are technical members (22%), software engineering process group member (22%) and just (4%) are team leaders.

Table 2. Respondent positions and experience.

| Position | Experience | | Total |
|---|---|---|---|
| | Less than 3 years | 3 -10 years | |
| Project or Team Leader | 1 (4%) | 5 (22%) | 6 (26%) |
| Manager | 0 (0%) | 1 (4%) | 1 (4%) |
| Technical Member | 5 (22%) | 2 (9%) | 7 (30%) |
| Software Engineering Process Group Member | 5 (22%) | 4 (17%) | 9 (39%) |
| Total | 11 (48%) | 12 (52%) | 23 (100%) |

### 4.2. Development and Measurement Issues

#### 4.2.1. Software Philosophy

In terms of what type of Software Philosophy that the organization follows when they develop web applications. Figure 2 indicates that majority of the respondent's use their own philosophy (44%), followed by using code and fix (30%), agile software development (22%) and waterfall (4%).which means that more than (70%) of respondents still not use any software development philosophy for developing web applications in their companies.

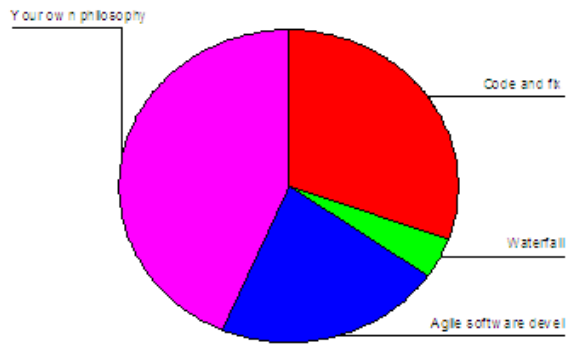

Figure 2. Software philosophy

#### 4.2.2. Development Methods That Respondents Are Familiar With

Regarding to the types of methodologies does the developers of small software firms are familiar with and have a good background about this section were developed in the questionnaire and were answer by the them based on that. Therefore, this part has been analyzed using multi response technique. Table 3 describes that the majority of respondents are familiar with Waterfall (74%) followed by XP (65%), Spiral model (30%), Scrum (22%), Prototyping (17%), Incremental (13%), AUP (13%), V-model (9%), DSDM (9%), FDD (4%) and lastly EUP (4%).

Table 3. Methodologies that respondents are familiar with

| Development Methods types | Frequency | Percent |
|---|---|---|
| Waterfall | 17 | 73.9 |
| V- Model | 2 | 8.7 |
| Spiral model | 7 | 30.4 |
| Agile Unified Process (AUP) | 3 | 13 |
| DSDM | 2 | 8.7 |
| FDD | 1 | 4.3 |
| Incremental | 3 | 13 |
| Prototyping | 4 | 17.4 |
| Enterprise Unified Process (EUP) | 1 | 4.3 |
| XP | 15 | 65.2 |
| Scrum | 5 | 21.7 |

**4.2.3. Development Methods That Respondents Are Familiar With**

Respondents were asked about the measurement types they use during the development and the type of the development methods that they currently used. The data was analyzed using cross tabulation analysis and multi response technique. Table 4 demonstrates that majority of the respondents are not using any specific type of measurements (65%) distributed according to the type of development methods as, no development method used (48%), Waterfall (9%), XP (4%) and Scrum (4%). Furthermore, (26%) of respondents are using use case point as type of measurement and distributed as, no development method used (9%), XP (4%), Waterfall (4%), Scrum (4%) and DSDM (4%). This mean that the majority of small software firms still not use any type of measurements while the majority of them still not use any systematic development method as well.

Table 4. Measurement type and development methods type

| Measurement type | Development method types | | | | | Total |
|---|---|---|---|---|---|---|
| | Waterfall | DSDM | XP | Scrum | No method | |
| Use Case Points | 1 (4.3%) | 1 (4.3%) | 1 (4.3%) | 1 (4.3%) | 2 (8.7%) | 6 (26.1%) |
| Function Points | 1 (4.3%) | 1 (4.3%) | 2 (8.7%) | 0 (0%) | 0 (0%) | 4 (17.4%) |
| Line of Code (LOC) | 0 (0%) | 0 (0%) | 0 (0%) | 0 (0%) | 1 (4.3%) | 1 (3.8%) |
| No specific type of measurement | 2 (8.7%) | 0 (0%) | 1 (4.3%) | 1 (4.3%) | 11 (47.8%) | 15 (65.2%) |
| Total | 4 (17.4%) | 1 (4.3%) | 4 (17.4) | 2 (8.7%) | 13 (56.5%) | 23 (100%) |

**4.2.4. Measurement Type and Measurement Methods**

Respondents were asked to indicate the type of software measurement they use and what measurement method which they apply to perform these measurements. Cross tabulation analysis was used for analyze the taken data, results for measurement type was attained using the multi response technique. According to Table 5 the majority of respondents do not use any measurement type (65.2%) and the majority of them still not use any methods for applying measurements. Whereas (22%) of respondents use PSM for performing the measurements process distributed according to the type measurement they use as, (17%) use the use case points, (13%) use the function points. Whilst, the percentage of respondents use the SPC method (9%) distributed according to the type measurement they use as, (4%) use function points (4%) and line of code. More over (4%) of respondents use the GQM all of them use the function points measurement type.

Table 5. Measurement types and methods types

| Measurement type | Measurement method | | | | Total |
|---|---|---|---|---|---|
| | GQM | PSM | SPC | No specific method used | |
| Use Case Points | 0 (0%) | 4 (17.4%) | 1 (4.3%) | 1 (4.3%) | 6 (26.1%) |
| Function Points | 1 (4.3%) | 3 (13%) | 0 (0%) | 0 (0%) | 4 (17.4%) |
| Line of Code (LOC) | 0 (0%) | 0 (0%) | 1 (4.3%) | 0 (0%) | 1 (4.3%) |
| No specific type of measurement | 0 (0%) | 0 (0%) | 0 (0%)) | 15 (65.2%) | 15 (65.2%) |
| Total | 1 (4.3%) | 5 (21.7%) | 2 (8.7%) | 16 (69.6%) | 23 (100%) |

**4.2.5. Measurement Stage and Development Method Type**

In this section respondents were asked about the stage of performing measurement within the development process and the type development method that were currently used. The data were analyzed by cross tabulation analysis. The results illustrates that that the majority of respondents do not use any specific measurements during the development (65%) distributed according to the development method used as, no development method used (48%), Waterfall (9%), XP (4%) and Scrum (4%). Furthermore, companies that prefer to perform measurement at the end of the coding phase (26%) distributed according to the development method used as, no method used (9%), using Waterfall (9%), using Scrum (4%) and DSDM (4%). Moreover, companies that prefer to use measurement early as soon as possible software projects were acquiring (9%) all of them are using XP. see Table 6. This mean the majority of respondents are not use measurements and the majority of them also still not use any specific development method.

Table 6. Measurement stage and development methods type

| Measurement stage | Development method type | | | | | Total |
|---|---|---|---|---|---|---|
| | Waterfall | DSDM | XP | Scrum | No method | |
| The end of the coding phase | 2 (8.7%) | 1 (4.3%) | 0 (0%) | 1 (4.3%) | 2 (8.7%) | 6 (26.1%) |
| Early as soon as possible software projects were acquiring | 0 (0%) | 0 (0%) | 2 (8.7%) | 0 (0%) | 0 (0%) | 2 (8.7%) |
| no measurement used | 2 (8.7%) | 0 (0%) | 1 (4.3%) | 1 (4.3%) | 11 (47.8%) | 15 65.2% |
| Total | 4 (17.4%) | 1 (4.3%) | 3 (13%) | 2 (8.7%) | 13 (56.5%) | 23 (100%) |

**4.2.6. Reasons of not using any development method**

In this part respondents of the survey were asked about the reason of not using and development method for building web applications. Data of this table calculated using multi response technique. Whereas the most of respondents answer that the current methodologies need specific training (90%) followed by the current methodologies need specific team to be performed (84%), also using the current methodologies take a lot of time (21%). However, only (5%) indicate that the current methodologies consume a lot of money. see Table 7.

Table 7. Reasons of not using the current dev method

| Reasons of Not Using the Current Methods | Frequency | Percent |
|---|---|---|
| Using any development method takes a lot of time | 4 | 21.1 |
| Consume a lot of money | 1 | 5.3 |
| Need specific team to be performed | 16 | 84.2 |
| Need specific training to be performed | 17 | 89.5 |

**4.2.7. Why Organizations do not Use Measurements**

This part aimed to identify the reasons of why company did not use any type software measurement with in the development process; so that respondents of small software firms in Jordan were asked to address these reasons. This part analyzed using the multi response technique. Respondents indicate that the majority of companies not aware of performing software measurements (72%) followed by software measurements need specific team to be performed (61%). Furthermore, (50 %) of respondents said that no body inside the company familiar with software measurements, consuming time reason takes (22%) and only (11%) of respondents said that using software measurement consume a lot of money. See Table 8.

Table 8. Why organizations do not use measurements

| Reasons of not using any specific measurements | Frequency | Percent |
|---|---|---|
| No body inside the company familiar with software measurement | 9 | 50 |
| Take a lot of time to employ software measurement | 4 | 22.2 |
| Consume a lot of money | 2 | 11.1 |
| Need specific team to perform | 11 | 61.1 |
| Your organization is not a ware to perform software measurement | 13 | 72 |

## 5. CONCLUSIONS

The pilot survey was conducted to check and validate the reliability and validity of the questionnaire to prepare the instruments and procedures which aim to modify the final questionnaire design. The objective of this study was to firstly determine the real characteristics of small software firms in Jordan. Secondly, examine the need of new methodology for developing web applications in small software firms. Thirdly, investigate and analyze the current web applications development and measurement practices for Jordanian small software firms. The findings showed that the majority of small software firms in Jordan have 10 to 30 employees followed by 31 to 50 employees which consistent with the finding of. Developers inside these firms have ten or less than ten years of experience and few managers and team leaders have more than ten years experience. In fact, the majority of respondents did not use any method that published in literature for developing web applications in small software firms which means there a need for new methodology for developing web applications in small software firms. Consequently, a great part of developers inside the targeted companies are familiar with Waterfall, Extreme programming (XP), Spiral and Scrum. However, respondents when they asked about the reason of not using specific development methods, a high percentage of them answered that using particular method need specific team to be performed and assume that when using specific method there is a need for team training.

On the other hand, The majority of respondents still not use any measurements on the development process whereas there is minimal percentage of them use function points, use case points and line of code after the coding phase, which means there is alack of deploy and perform measurements types and methods with in the development process. Consequently, respondents when asked about why they are not using any specific measurements or method the majority of them explain that because no body inside the company familiar with measurements type and methods and also using specific measurement need specific trained team to be performed. Based on the above, it is clearly obvious that there is alack of performing and applying the important measurement and development practice within the development methods that currently used by small software firms in Jordan. Therefore, the findings of the pilot study will be used for building methodology for developing web applications in small software firms which integrated with the important measurement and development practices to get a high quality product. The successful execution of this pilot study signified that the instrument of questionnaire and analysis are valid and reliable to be used for the actual survey.